\documentclass[useAMS]{mn2e}

\usepackage{amssymb,amsmath,psfig,times}
\usepackage{graphicx}
\def\gsim{ \lower .75ex \hbox{$\sim$} \llap{\raise .27ex \hbox{$>$}} }
\def\lsim{ \lower .75ex\hbox{$\sim$} \llap{\raise .27ex \hbox{$<$}} }

\def\sc{Schwarzschild}
\def\beq{\begin{equation}}
\def\eeq{\end{equation}}

\def\sc{Schwarzschild}

\def\sw{{\it Swift}}
\def\fe{{\it Fermi}}

\def\ep{$E_{\rm peak}$}
\def\epf{$E_{\rm peak}-P$}

\def\pf{$P$}

\def\ama{$E_{\rm peak}-E_{\rm iso}$}
\def\yone{$E_{\rm peak}-L_{\rm iso}$}

\title[Fermi Short GRBs]{Spectral evolution of {\it Fermi}/GBM short Gamma--Ray Bursts}
\author[G. Ghirlanda et al.]{G. Ghirlanda
$^{1}$\thanks{E-mail:giancarlo.ghirlanda@brera.inaf.it}, G. Ghisellini$^{1}$,
L. Nava$^{2}$, D. Burlon$^{3}$\\
$^{1}$INAF -- Osservatorio Astronomico di Brera, Via E. Bianchi 46, I-23807 Merate, Italy\\
$^{2}$SISSA -- via Bonomea, 265, I-34136 Trieste, Italy\\
$^{3}$Max-Planck-Institut f\"{u}r Extraterrestrische Physik, Giessenbachstra\ss e 1, 
D-85478 Garching, Germany}
\begin{document}

\date{}


\maketitle

\label{firstpage}

\begin{abstract}
We study the spectral evolution of 13 short duration Gamma Ray Bursts (GRBs) detected by 
the Gamma Burst Monitor (GBM) on board \fe.
We study spectra resolved in time at the level of 2 -- 512 ms 
in the 8 keV--35 MeV energy range.
We find a strong correlation between the observed 
peak energy \ep\ and the flux \pf\ within individual short GRBs. 
The slope of the $E_{\rm peak}\propto P^s$
correlation for individual bursts ranges between $\sim$0.4 and $\sim$1.
There is no correlation between the low energy spectral index and the peak energy 
or the flux.  
Our results show that in our 13 short GRBs
\ep\ evolves in time tracking the flux. 
This behavior is similar to what found in the population of long GRBs and it is in 
agreement with the evidence that long GRBs and (the still few) 
short GRBs with measured redshifts follow the same rest frame \yone\ correlation. 
Its origin is most likely to be found in the radiative mechanism that has to be the 
same in both classes of GRBs.  
\end{abstract}


\section{Introduction}

Short Gamma Ray Bursts have been 
a challenge since the finding of their spectral diversity with 
respect to the class of long GRBs (e.g. see Nakar 2007; Lee \& Ramirez--Ruiz 2007 
for recent reviews). 
Short GRBs have optical and X--ray afterglows, and in few cases 
they show also X--ray flares, similar to those of long bursts but scaled 
by their fluence (e.g. Gehrels et al. 2008; Nysewander et al. 2008). 
However, the class of short GRBs is somewhat heterogeneous
for what concerns the prompt emission properties (e.g. there are short 
GRBs followed by a faint extended emission -- Norris \& Bonnell 2006; 
Donaghy et al. 2006; Norris \& Gehrels 2008) or 
the host galaxies properties (short GRBs are found in almost all 
galaxy types -- e.g.  Berger 2009). 
Recently, this picture has been also complicated by the detection of long GRBs 
at very high redshifts like GRB 080913 at $z=6.7$ (Greiner et al. 2009), and  GRB 090423 at $z\simeq8.2$ (Salvaterra et al. 2009, Tanvir et al. 2009) which have an
intrinsic duration of less than 2 s.
Classification schemes of short GRBs that try to merge all these
evidences have been proposed (Zhang et al. 2007). 

For what concerns the prompt emission, it was discovered through the BATSE 
sample that short GRBs are spectrally harder than long 
bursts (Kouveliotou et al. 1993). 
Detailed analysis of the time integrated spectra of short GRBs 
(Paciesas et al. 2003, Ghirlanda, Ghisellini \& Celotti 2004, Ghirlanda et al. 2009)  
have also shown that this spectral difference is due to a harder low 
energy spectral component in short GRBs (but see Nava et al. 2010). 
These studies also revealed that, on average, the first 2 seconds of emission 
of long GRBs have similar spectral properties of short bursts 
(as also found from the comparison of the variability patterns 
-- Nakar \& Piran 2002), thus suggesting the presence of a common emission mechanism 
that operates in these two kind of sources. 

Considered individually, long GRBs show a strong spectral evolution with two 
possible behaviors: the peak energy \ep\ of the 
$\nu F_{\nu}$ spectrum decays in time (``hard to soft" evolution) or 
follows the variation of the flux (``tracking" evolution;
e.g. Band et al. 1993; Ford et al. 1995). 
This emerged from the time resolved spectral analysis of long GRBs observed by BATSE. 
Recently Lu, Hou \& Liang (2010) reported that the
hard to soft spectral evolution is dominant in long bursts
(at least initially, for 2/3 of their sample of 22 single pulses BATSE GRBs). 
To date, however, no detailed study of the possible spectral evolution of short GRBs was performed. 

Another recent issue concerning the prompt emission of GRBs is the nature 
of the spectral--energy correlations discovered for the sample of long GRBs with 
measured redshifts. Amati et al. (2002) found that the rest frame peak energy correlates with the 
isotropic energy (the \ama\ correlation) while Yonetoku et al. (2004) showed 
that a similar correlation exists between \ep\ and the isotropic luminosity 
(the \yone\ correlation). 
The open debate is if these correlations have a physical foundation 
(Ghirlanda et al. 2005; Bosnjak et al. 2008; Ghirlanda et al. 2008; 
Nava et al. 2008; Krimm et al. 2009; Amati et al. 2009) or if they are 
the result of selection effects (Nakar \& Piran 2005; Band \& Preece 2005; 
Butler et al. 2007; Butler et al. 2009; Shahmoradi \& Nemiroff 2009).  
Both the \ama\ and \yone\ correlations are defined by considering the 
time integrated spectral properties of long GRBs with measured redshifts. 
Recently, Ghirlanda et al. (2009) showed that also short GRBs with measured 
redshifts follow the same \yone\ correlation defined by long bursts, but 
not the corresponding \ama\ correlation. 

One of the most stringent result supporting a physical origin of the 
\yone\ correlation is that, in long GRBs with measured redshifts detected by 
\fe, there is a strong \yone\ correlation {\it within individual bursts} (Ghirlanda et al. 2010)
which cannot be due to selection effects but must have a physical origin.
A similar result was reached by Firmani et al. 2009 based on the time resolves
spectral analysis of long GRBs detected by \sw. 
Several interpretations for the \yone\ (or \ama) correlations 
have been proposed (e.g. Yamazaki et al. 2004; Lamb et al. 2005; 
Rees \& Meszaros 2005; Levinson \& Eichler 2005; Toma et al. 2005; 
Eichler \& Levinson 2005; Barbiellini et al. 2006; Thompson 2006; 
Ryde et al. 2006; Giannios \& Spruit 2007; Thompson et al. 2007; 
Guida et al. 2008; Panaitescu 2009), but there is no general consensus about
a prevalent idea.

Therefore, the two topics we aim to address in the present work are (1) 
whether, and how, the spectra of short GRBs evolve in time, (2) if also 
in short GRBs there is a spectral--energy correlation between the 
peak energy and flux similar to what found in long events. 

Time resolved spectral studies of short GRBs are hampered by their lower 
fluence and duration with respect to long GRBs. 
Moreover, time resolved 
spectroscopy requires an instrument with the largest energy coverage and 
good spectral resolution in the keV--MeV energy range in order to constrain 
the spectral parameters of individual (time--resolved) spectra and to follow 
their temporal variation. 
\sw\ detected a large number of short GRBs, but it is not suited to 
this aim because of the narrow energy range of the BAT instrument (15--150 keV). 
The Gamma Burst Monitor (GBM; 8 keV--40 MeV) on--board the \fe\ 
satellite, instead, represents a unique opportunity to 
study, in details for the first time, how the spectrum in short GRBs evolves with time. 

In \S2 we present the sample of short \fe\ GRBs selected for the time 
resolved spectral analysis. The details of the spectral analysis are described 
in \S3 and the results in \S4. We draw our conclusions in \S5.

\section{The sample}

We consider the 237 GRBs detected by the \fe\ GBM up to May 2010 and whose detection 
has been published through GCN communications. 
In this sample there are 37 short GRBs 
with observed duration (as reported in the GCNs) $\le$2s. 
For the time resolved spectral analysis 
we consider the short GRBs with the 
largest fluence and peak flux.
This ensures to have, for each GRB, a set of time resolved 
spectra with enough signal to constrain the spectral parameters. 
We select the 14 short GRBs with 
a fluence larger than $8\times 10^{-7}$ erg cm$^{-2}$ 
(integrated in the 8 keV -- 1 MeV energy range)
and a peak photon flux  $P_{\rm peak}\ge 11$ ph cm$^{-2}$ s$^{-1}$.

We anticipate that for one burst (GRB 100223) the detector response 
files are not present in the archive yet so that its spectral analysis was not possible.  
GRB 081113 has a time integrated spectrum which is consistent with a 
single power law (von Kienlin et al. 2008a) and similarly its time resolved spectra are all consistent 
with a single power law. All the other GRBs of our sample have time integrated spectra (as reported in the 
GCN) with a peak in $\nu F_{\nu}$. 
We have included GRB 090308 and GRB 081107 in the sample although their duration ($T_{90}$)
is somewhat longer than 2 s (i.e. 2.2 s and 2.11 s respectively), 
since they have a fluence and peak flux larger than our 
threshold and considering that the division of short and long GRBs at 2 s is not sharp, 
since both populations have 
a gaussian $T_{90}$ distribution extending below and above this time cut. 
These two bursts are reported separately in Tab. \ref{tab1}.
Therefore, the sample of analyzed short GRBs is composed by 14 events. 

In Tab. \ref{tab1} we report the names, duration, fluence and peak flux of the selected short GRBs 
(with the corresponding reference -- Col. 4). 
The redshift is not measured for all the bursts in our sample 
except for GRB 090510 at $z=0.903$ (Rau et al. 2009). 

\begin{table}  
\caption{Short GRB sample. The last three digits in parenthesis are the 
fractional trigger number assigned to the GRB by the GBM archive. 
The fluence $F_{-6}$ is in units of 10$^{-6}$ erg cm$^{-2}$.
References for the duration ($T_{90}$), fluence $F_{-6}$ and peak flux $P_{\rm peak}$: 
(1) Guiriec et al. 2009; 
(2) Guiriec et al. 2009a; 
(3) von Kienlin et al. 2009; 
(4) McBreen et al., 2008; 
(5) Wilson-Hodge et al., 2009; 
(6) Bissaldi et al., 2009; 
(7) Goldstein et al 2009; 
(8) Bissaldi et al., 2008; 
(9) Êvon Kienlin et al., 2008; 
(10) Bissaldi et al., 2009a; 
(11) Goldstein et al 2009a; 
(12) von Kienlin et al., 2010; 
(13) Goldstein et al., 2009b; 
(14) von Kienlin et al., 2008a. 
Col. 6 lists the range of time resolution in milliseconds used to perform the spectral analysis. 
In the case of GRB 090228 the time resolution is 2 ms for the first peak and 32 ms for the (fainter) 
second one of its lightcurve. 
$^{\mathrm{*}}$ GRBs with  a precursor. GRB~090308 and GRB~081107 are shown separately since they 
have a slightly larger duration 
than the canonical 2s dividing line between short and long GRBs but still have a fluence and 
peak flux above our selection thresholds. }
\label{tab1} 
\begin{center}
\begin{tabular}{l|lllll}
\hline
\noalign{\smallskip}
 ÊGRB &$T_{90}$ &$F_{-6}$ Ê Ê Ê Ê Ê Ê& $P_{\rm{peak}}$ Ê Ê Ê Ê Ê &Ref Ê&$T_{\rm res}$ Ê\\
 Ê Ê Ê Ê Ê Ê Ê Ê&s Ê Ê Ê & erg cm$^{-2}$ Ê Ê Ê& ph cm$^{-2}$ s$^{-1}$ Ê& Ê Ê & Êms Ê Ê Ê Ê Ê\\
 \noalign{\smallskip}
\hline   
\noalign{\smallskip}
  090510(016)$^{*}$ & 1.0 Ê & 30.4$\pm$2.0 Ê Ê& Ê80 Ê Ê Ê Ê Ê Ê		&1 Ê& 16--64 Ê\\
 Ê090227(772) & 0.9 Ê & 8.7$\pm$0.1 Ê Ê & Ê34.6$\pm$0.3 Ê	&2 Ê& 8--32 Ê\\
 Ê090228(204) & 0.8 Ê & 6.1$\pm$0.09 Ê Ê& Ê133$\pm$8 Ê Ê 	&3 Ê& 2--32 \\
 Ê081216(531)$^{*}$ & 0.96 Ê& 3.6$\pm$0.1 Ê& Ê55$\pm$3 Ê Ê Ê	&4 Ê& 16--32 \\
 Ê090305(052) & 2.0 Ê & 2.7$\pm$0.2 Ê Ê & Ê11$\pm$2 Ê Ê Ê	&5 Ê& 64--0.448 \\
 Ê090902(401) & 1.2 Ê & 2.11$\pm$0.14 Ê & Ê11.4$\pm$1.3 Ê	&6 Ê& 64--0.512 \\
 Ê090108(020) & 0.9 Ê & 1.28$\pm$0.24 Ê & Ê39.7$\pm$3.9 Ê	&7 Ê& 16--0.384 \\
 Ê081223(419) & 0.89 Ê& 1.2$\pm$0.10 Ê Ê& Ê22$\pm$3 Ê Ê Ê	&8 Ê& 64--0.704 \\
 Ê081113(230) & 0.5 Ê & 1.07$\pm$0.03 Ê & Ê20$\pm$1 Ê Ê Ê	&9 Ê& ... \\
 Ê090206(620) & 0.8 Ê & 1.04$\pm$0.06 Ê & Ê19$\pm$1 Ê Ê Ê	&10 & 64 \\
 Ê090328(713) & 0.32 Ê& 0.96$\pm$0.03 Ê & Ê29.83$\pm$2.38&11 & 32--64 \\
 Ê100206(563) & 0.13 Ê& 0.93$\pm$0.04 Ê & Ê31$\pm$2 Ê Ê Ê	&12 & 32Ê\\
  \hline
 Ê090308(734) & 2.11 Ê& 3.46$\pm$0.13 Ê & Ê14.22$\pm$0.91&13 & 64 \\
 Ê081107(321) & 2.2 Ê & 1.64$\pm$0.28 Ê & Ê11$\pm$3 Ê Ê Ê&14 & 64 \\
  \noalign{\smallskip}
\hline
\end{tabular}
\end{center}
\end{table}

\section{Spectral analysis}
\begin{figure*}
\psfig{file=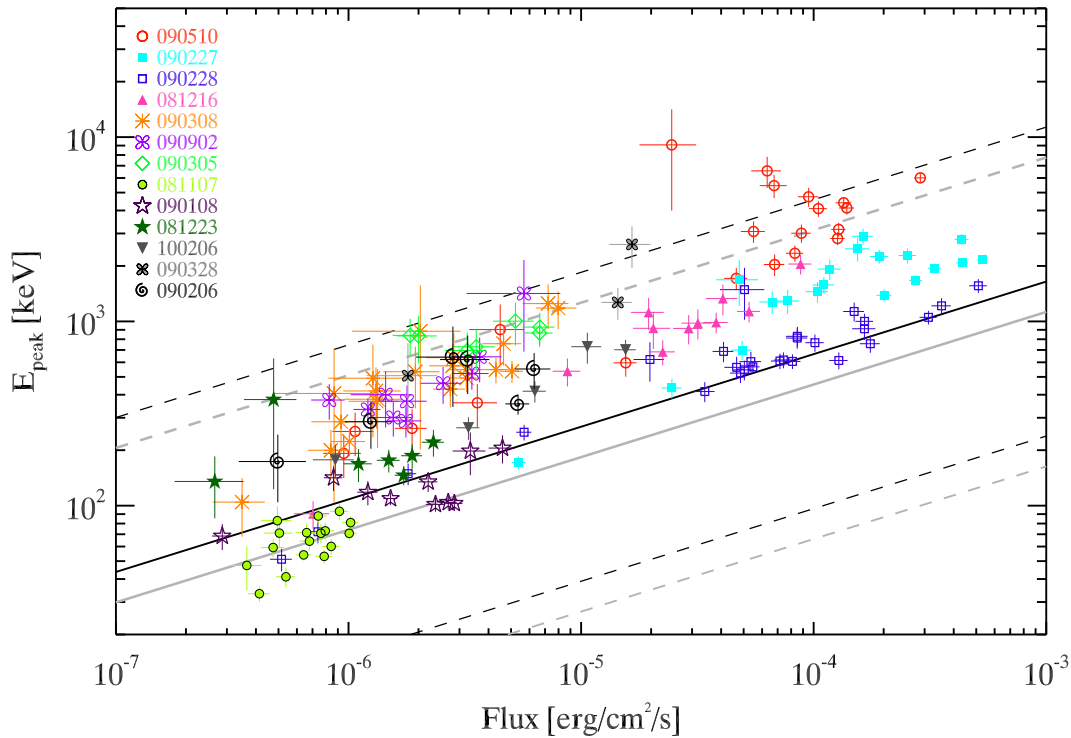,width=16cm,height=11cm}
\caption{
Correlation between the peak energy \ep\ and the flux of the 163 time resolved spectra of the 13 
short GRBs analyzed in this work (Tab. \ref{tab1}). Different symbols/colors 
correspond to different bursts (as shown in the legend). 
The solid line (dashed lines) is the ``Yonetoku" relation (and its 3$\sigma$ scatter) of long GRBs with 
measured redshifts (adapted from Ghirlanda et al. 2009) 
transformed in the observer frame assuming $z=1$. The grey (solid and dashed lines) are for $z=0.5$.
}
\label{fg1}
\end{figure*}

The GBM (Meegan et al. 2009) comprises 12 thallium 
sodium iodide [NaI(Tl)] and two bismuth 
germanate (BGO) scintillation detectors which cover the energy ranges  
$\sim$8 keV--1 MeV and $\sim$300 keV--40 MeV, respectively.
The GBM acquires three types of data suited for spectroscopic studies (Meegan et al. 2009): 
the ``CTIME" data consisting of a sequence spectra (binned in 8 energy channels) 
with a time resolution between 
0.256 and 1.024
seconds, the ``CSPEC" data  containing a 
sequence of 128 energy channel spectra binned in time with a variable resolution between 
1.024 s and 4.096
s and the ``TTE" event data files containing individual photons with 
time and energy tags. 

To the aim of studying short GRBs, TTE data are ideal because they allow to choose the 
temporal resolution according to the burst duration and flux. Due to the limited buffer size, 
TTE data only covers the interval between $\sim$30 s before and $\sim$300 s after the burst trigger time. 
This time interval fully encompasses the duration 
of short GRBs and allows to fit the background spectrum by selecting time intervals 
before and after the burst. 

For the time resolved spectral analysis we used the recently released software 
RMFIT\footnote{http://fermi.gsfc.nasa.gov/ssc/data/analysis/user/} (\texttt{v33pr7}). 
In order to model the background spectrum for the time resolved spectroscopic analysis, we 
selected two time intervals before and after the burst. The sequence of background spectra in the 
two selected intervals were fitted with a first order polynomial to account for 
the possible time variation of the background spectrum. 
Then the background spectrum was extrapolated to the time 
intervals selected for the time resolved spectroscopy of individual GRBs. 

For each burst, we jointly fitted the spectra from the NaI detectors which 
had the largest illumination by the GRB.  
The NaI detectors were selected for having an angle of position of the GRB with 
respect to the detector normal lower than $\sim$80 degrees. Among the two BGO detectors
we selected that with the smallest angle to the GRB. 
Four NaI detectors and one BGO were analyzed for GRB 100206, 090902, 081213, 090108, 090305,
090308, 081216, 090228, 090227B; three NaI and one BGO for GRB 090206, 090328 and three 
NaI for GRB 081107. For GRB 090510 we used 5 NaI and both BGO data.
The inclusion of the BGO data extends the spectral coverage of the NaI detectors 
from 1 MeV to $\sim$35 MeV. 
This energy extension represents an unprecedented opportunity 
for the spectral analysis of short \fe\ GRBs whose time integrated spectrum has typically a 
peak energy of the order of 1 MeV (e.g. Nava et al. 2010). 

For each burst we performed a time resolved spectral analysis by opportunely changing the 
time resolution  
(starting with 64 ms time resolution of the TTE data) and we checked that each 
time resolved spectrum had well constrained spectral parameters (i.e. the normalization, \ep\ and $\alpha$
are required to have relative errors less than 100\%).
For the brightest GRBs of our sample (i.e. GRB 090227B, 090228) we 
performed a time resolved spectral analysis down to the 
8 ms and 2 ms timescale. 
In some cases, the end of the slow decline of the light curves did not ensure enough signal to 
obtain a good fit so that a coarser time resolution was applied (e.g. a single bin of 
0.5 s at the end of GRB 090902 was analyzed).
In Tab. \ref{tab1} (last column) we report the minimum and maximum
time resolution in milliseconds at which the time resolved spectral 
analysis was performed.

The model adopted is a power law with an exponential cutoff whose 
free parameters are the low energy spectral index $\alpha$, the peak energy 
$E_{\rm peak}$ (i.e the peak of the $\nu F_{\nu}$ spectrum), and the normalization. 
We also allowed for a variable normalization factor in order to fit together the 
data of the NaI and BGO detectors. 
For the time resolved spectral analysis we fixed the normalization factors to the 
values obtained from the fit of the time integrated spectra of each burst. The values of these
normalization factors are all consistent with 1, with a deviation  
of 20\% at most.
The choice of this model is motivated by the fact that at high energies the 
response of the BGO rapidly decreases for increasing energy so that it is hard, 
in single time resolved spectra,  to constrain the possible presence of a power 
law component of e.g. the Band function (Band et al. 1993) which is instead typically 
fitted to the time integrated spectra of long GRBs. 
We tried to fit the time resolved spectra with the Band function but in most cases the high energy spectral
index $\beta$ was unconstrained. This is confirmed also by the 
recent analysis of the brightest three short GRBs of our sample 
(Guiriec et al. 2010) that shows that the fit with 
a Band function results in unconstrained $\beta$ in most 
of their time resolved spectra.
Therefore, for homogeneity and with the 
aim of comparing the spectral evolution trends in short GRBs, we 
adopted the same spectral model, i.e. a cutoff--power law for 
all the time resolved spectra.

 

\begin{figure}
\vskip -0.2cm
\hskip 0.3cm
\psfig{file=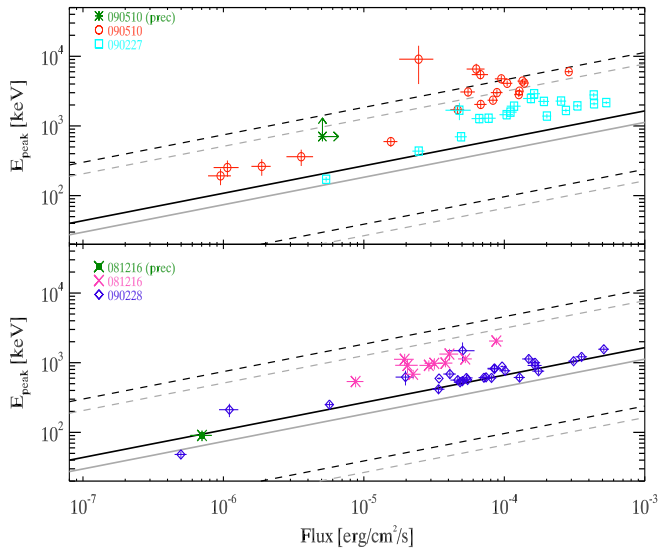,width=8.5cm,height=7cm}
\vskip0.4cm
\caption{
Correlation between \ep\ and the flux $P$ for the 4 GRB with the highest fluence. 
The upper panel shows GRB 090510 and GRB 090227B. 
The precursor of GRB 090510 is shown with the green arrow symbols since its spectrum is consistent with a single powerlaw. 
The lower panel shows 
GRB 081216 (with a precursor, green symbols) and GRB 090228. 
Color code (data points, solid and dashed lines) as in Fig. \ref{fg1}. 
}
\label{panel}
\end{figure}

\section{Results}

\begin{figure}
\vskip -0.4 cm
\hskip -0.8cm
\psfig{file=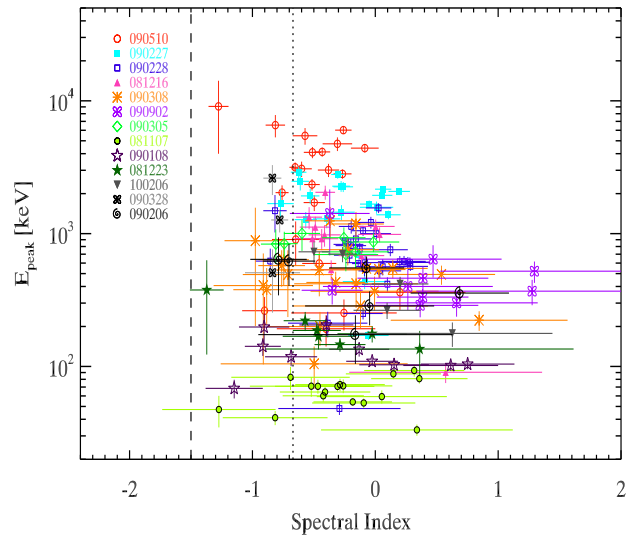,width=9.5cm,height=8cm}
\vskip -0.4 cm
\caption{
\ep\ versus the photon spectral index of the fitted model (cutoff--power law). 
Same symbols and colors as in Fig. \ref{fg1}. 
The vertical lines represent the limits predicted by the standard synchrotron model 
(dotted line) and in the case of cooling (dashed line).
}
\label{fg2}
\end{figure}

We searched for possible correlations between the best fit spectral 
parameters of the time resolved spectra of our sample of short GRBs. 
We find that there is a strong correlation between the peak energy \ep\ 
and the flux \pf\ within individual short GRBs. The flux \pf\ is calculated by integrating 
the best fit model of each time resolved spectrum in the 8 keV--35 MeV energy range. 
This correlation is shown in Fig. \ref{fg1} for the whole sample of 13 GRBs. We 
could extract for them a total of 163 time resolved spectra. 
Considering the 13 short 
GRBs individually, the  \ep--\pf$^s$ correlation of individual GRBs has a slope 
that ranges between 0.4 and 1.0. 
The \epf\ correlation is in the observer frame (only GRB 090510 has a known redshift). 
For comparison with the population of long GRBs, we 
show in Fig. \ref{fg1} the \yone\  (``Yonetoku") correlation of long GRBs with measured redshifts 
(adapted from Ghirlanda et al. 2010) assuming a redshift $z=1$  
and $z=0.5$ (solid black and grey lines respectively). 
Also shown are the boundaries representing the 3$\sigma$ scatter
of this correlation (dashed lines in Fig. \ref{fg1}). 
The short GRBs spectral evolution trends are consistent 
with this correlation (transformed in the observed frame) as already shown for the spectral evolution 
tracks of long GRBs (Ghirlanda et al. 2010). 
However, we note from Fig. \ref{fg1} that the \epf\ correlation defined by short GRBs is similar 
in slope to the \yone\ correlation transformed in the observer frame (solid grey and black lines in 
Fig. \ref{fg1}) but it has a higher normalization.

The most sampled events (i.e. those studied on the shortest integration timescales -- 
down to 4 ms in the case of the peak of GRB 090228) are the 4 GRBs with the largest 
fluence/peak flux (Tab. \ref{tab1}),  
shown separately in Fig. \ref{panel}.  
Noteworthy, two of them also show precursor activity which is shown 
with different symbols  (colors) in Fig.\ref{panel}. 
GRBs are known to show, in $\sim15$\% of cases 
(Lazzati 2005, Burlon et al. 2008), emission preceding the main episode (i.e. precursors). 
Burlon et al. (2009) analyzed a large sample of bright long BATSE GRBs with precursor emission 
and showed that the same spectral evolution is present in the precursors and in the main GRB events.
They also did not find any relation between the spectral index and \ep\, 
similarly to what shown in Fig. \ref{fg2}.
Here we find that, although still based only on two cases, also in short GRBs the precursors have spectra which are consistent with 
the general trend of the main event.
This result, shown for short GRBs in this paper for the first time, points towards a common origin 
of the precursor and the main emission.

Fig. \ref{fg2} shows that there is no correlation between the low energy spectral 
index $\alpha$ of the cutoff--power law model and the peak energy \ep. 
Similarly we do not find any correlation between $\alpha$ and the flux \pf. 
We instead find, as already shown by the analysis of time integrated spectra of short 
GRBs (e.g. Ghirlanda et al. 2009), that they are harder than long ones 
and this makes a large fraction of them inconsistent with the ``lines of death" of 
synchrotron emission (with no cooling -- vertical dotted line in Fig. \ref{fg2}) 
and makes all of them inconsistent with synchrotron emission
by cooling electrons (dashed vertical line in Fig. \ref{fg2}).


\section{Conclusions}

The time resolved spectra of individual short GRBs evolve in time, 
and their \ep\ is strongly correlated with their flux.
Furthermore, the found correlation is very similar to what already found
in individual long GRBs. 
The GBM data for the brightest bursts allowed to analyze their spectra even at the 
2--8 ms time resolution for part of the duration of these GRBs.
With respect to typical peak fluxes measured on $\sim$1 s time bins, 
the peak fluxes reached by our short GRBs, measured on a fraction of second, are extreme, 
i.e. of the order of several$\times 10^{-4}$ erg cm$^{-2}$ s$^{-1}$
(see Fig. \ref{fg1}). 

We do not know if these flux levels are reached also in long GRBs, since a time
resolved analysis with this degree of accuracy has not yet been done
(but time resolved spectra with a coarser time resolution for the 
brightest long bursts had much smaller peak fluxes, e.g. Kaneko et al. 2006).
The fact that the \epf\ correlation is similar suggests that 
long and short GRBs share the same emission mechanism for their prompt emission.
Furthermore, the emission mechanism should not depend on the progenitor,
(e.g. fireball--funnel interactions) if long and short GRBs do have different progenitors. 

For the brightest bursts, the dynamic range of the \epf\ is very large, being
more than 2 orders of magnitude in flux and more than one in \ep.
Even precursors obey the same \epf\ correlation, although only two short GRBs in our 
sample show a precursor.

The three brightest short bursts in our sample reach values of \ep\ significantly larger than those 
of long GRBs (e.g. Ghirlanda et al. 2010). On the other hand, they also reach {\it significantly larger fluxes} 
while remaining on the same \epf\ correlation defined by long bursts. For the same fluxes they do have the same 
\ep.

The \epf\ correlation, as well as the analogous Yonetoku relation for time integrated
spectra of different GRBs, has not yet received a convincing and broadly accepted
explanation, as mentioned in the introduction.
The fact that also short bursts obey it makes the search for a convincing 
interpretation even more compelling.

\section*{Acknowledgments}
We thank  the referee for useful comments and suggestions.
We acknowledge ASI (I/088/06/0) and a 2010 PRIN--INAF grant for
financial support. DB is supported through DLR 50 OR 0405. 
This research has made use of the data obtained through
the High Energy Astrophysics Science Archive Research Center Online Service,
provided by the NASA/Goddard Space Flight Center. DB and LN thank  
the Brera Observatory for the kind hospitality during the completion of this work.

\end{document}